\documentclass[aps,twocolumn,showpacs,amsmath,amssymb,prl]{revtex4-1}

\usepackage{epsfig}
\usepackage{graphicx}
\usepackage{booktabs}
\usepackage{multirow}
\usepackage{tabularx}
\usepackage{times}
\usepackage{color}
\usepackage{subfigure}
\usepackage{todonotes}
\usepackage[english]{babel}

\definecolor{Red}{rgb}{1,0,0}

\definecolor{Blu}{rgb}{0,0,01}

\definecolor{Green}{rgb}{0,1,0}

\newcommand{\be}{\begin{equation}}
\newcommand{\ee}{\end{equation}}

\newcommand\varpm{\mathbin{\vcenter{\hbox{%
  \oalign{\hfil$\scriptstyle+$\hfil\cr
          \noalign{\kern-.3ex}
          $\scriptscriptstyle({-})$\cr}%
}}}}
\newcommand\varmp{\mathbin{\vcenter{\hbox{%
  \oalign{\hfil$\scriptstyle-$\hfil\cr
          \noalign{\kern-.3ex}
          $\scriptscriptstyle({+})$\cr}%
}}}}

\newcommand{\vecn}{\boldsymbol{n}}
\newcommand{\vecm}{\boldsymbol{m}}

\newcommand{\vecx}{\boldsymbol{x}}

\begin{document}

\title{Controlling the Superconducting Transition by Rotation of an Inversion Symmetry-Breaking Axis}

\author{Lina G. Johnsen}

\affiliation{Center for Quantum Spintronics, Department of Physics, Norwegian
University of Science and Technology, NO-7491 Trondheim, Norway}

\author{Kristian Svalland}

\affiliation{Center for Quantum Spintronics, Department of Physics, Norwegian
University of Science and Technology, NO-7491 Trondheim, Norway}

\author{Jacob Linder}

\affiliation{Center for Quantum Spintronics, Department of Physics, Norwegian
University of Science and Technology, NO-7491 Trondheim, Norway}

\begin{abstract}

We consider a hybrid structure where a material with Rashba-like spin-orbit coupling is proximity coupled to a conventional superconductor. We find that the superconducting critical temperature $T_c$ can be tuned by rotating the vector $\vecn$ characterizing the axis of broken inversion symmetry. This is explained by a leakage of $s$-wave singlet Cooper pairs out of the superconducting region, and by conversion of $s$-wave singlets into other types of correlations, among these $s$-wave odd-frequency pairs robust to impurity scattering. These results demonstrate a conceptually different way of tuning $T_c$ compared to the previously studied variation of $T_c$ in magnetic hybrids.

\end{abstract}

\date{\today}

\maketitle

\textit{Introduction}.--- Over the last years, research on combining superconducting and magnetic materials has shown that the physical properties of the resulting hybrid structure may be drastically altered compared to those of the individual materials \cite{linder_np_15,eschrig_pt_11,eschrig_rpp_15}. In a conventional superconductor (S), electrons combine into $s$-wave singlet Cooper pairs \cite{bardeen_pr_57}. A decrease in the $s$-wave singlet amplitude leads to a loss of superconducting condensation energy, and thus also a suppression of the superconducting critical temperature, $T_c$. Such a decrease can be obtained by leakage of Cooper pairs into a nonsuperconducting material in proximity to the superconductor, and by conversion of $s$-wave singlets into different singlet and triplet Cooper pairs. For the latter to happen, the nonsuperconducting material must introduce additional symmetry-breaking. This is the case in superconductor-ferromagnet hybrids where the spin splitting of the energy bands of the homogeneous ferromagnetic material (F) leads to creation of opposite-spin triplets \cite{eschrig_pt_11,eschrig_rpp_15,linder_rmp_19}. 

A single, homogeneous ferromagnet cannot alone cause variation in the $s$-wave singlet amplitude under rotations of the magnetization $\vecm$. However, experiments \cite{gu_prl_02, moraru_prl_06, leksin_prl_12, banerjee_natcom_14, wang_prb_14} have demonstrated that the critical temperature of F/S/F and S/F/F structures can be modulated by changing the relative orientation of the magnetization of the ferromagnets. The misalignment opens all three triplet channels, leading to a stronger decrease in the superconducting condensation energy associated with the singlet amplitude.
Recent work \cite{jacobsen_prb_15, ouassou_sr_16,simensen_prb_18, banerjee_prb_18, johnsen_prb_19} has shown that the rotational invariance of the S/F structure can also be broken by adding thin heavy normal metal layers that boost the interfacial Rashba spin-orbit coupling. Spin-orbit coupling (SOC) introduces inversion symmetry-breaking perpendicular to an axis, here characterized by the vector $\vecn$.

\begin{figure}
    \centering
    \includegraphics[width=\columnwidth]{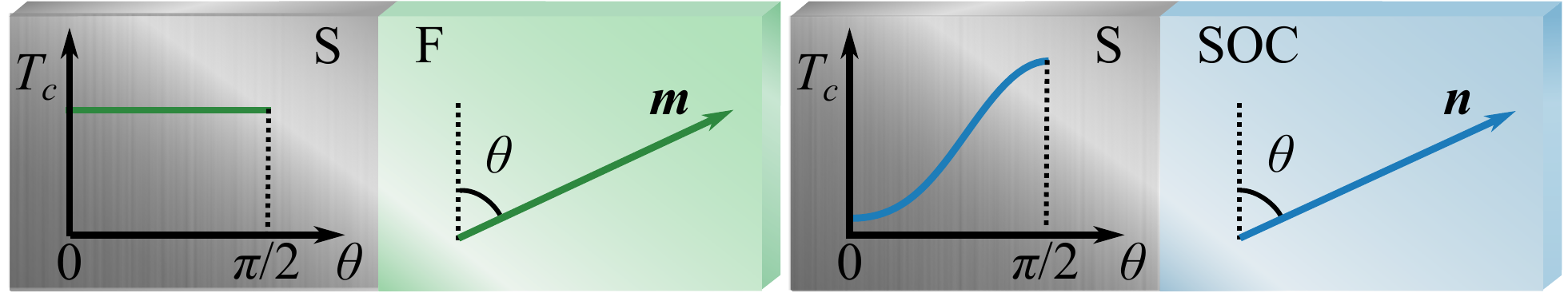}
    \caption{In a S/F bilayer (left), $T_c$ is invariant under a rotation of $\vecm$. In a S/SOC bilayer (right), the inversion symmetry perpendicular to $\vecn$ is broken. This opens up the possibility for a variation in $T_c$ under a rotation of $\vecn$.}
    \label{fig:TcFSOC}
\end{figure}

While ferromagnetism only leads to spin splitting of the energy bands of spin-up and spin-down electrons, Rashba SOC is in addition odd under inversion of the momentum component perpendicular to $\boldsymbol{n}$. This raises an interesting question. While the proximity effect and accompanying change in $T_c$ in a S/F bilayer is invariant under rotations of $\vecm$, is it possible that $T_c$ in a S/SOC bilayer is \textit{not} invariant under rotations of $\vecn$ (see Fig.~\ref{fig:TcFSOC})?

Motivated by this, we explore the possibility of $T_c$ modulation under reorientations of the inversion symmetry-breaking vector $\vecn$ in a bilayer consisting of a conventional superconductor and a material with Rashba-like SOC in the bulk. We also include interfacial Rashba SOC with an inversion symmetry-breaking vector $\vecn_{\text{int}}$ perpendicular to the interface. This simple model illustrates the concept of tuning $T_c$ via rotation of $\vecn$.

When the bulk SOC is stronger than than the interfacial contribution, we discover a suppression of $T_c$ when rotating $\vecn$ from an out-of-plane (OOP) to an in-plane (IP) orientation. This effect is enhanced by increasing the interfacial SOC, provided that $\vecn||\vecn_{\text{int}}$ when $\vecn$ is OOP. The difference in $T_c$ for IP and OOP orientations of $\vecn$ can at least partly be accounted for by the absence of $s$-wave odd-frequency triplets for an OOP orientation of $\vecn$. Since $s$-wave triplets are robust with respect to impurity scattering, we expect our prediction of an IP suppression of $T_c$ to be observable not only in the ballistic limit covered by our theoretical framework, but also in the diffusive limit. 
When interfacial SOC dominates, the $T_c$ modulation changes qualitatively. The critical temperature is instead suppressed for anti parallel compared to parallel $\vecn$ and $\vecn_{\text{int}}$. This is explained by a reduced leakage of $s$-wave singlets into the nonsuperconducting region when the total SOC magnitude is increased.
Moreover, we demonstrate a variation in $T_c$ even when $\vecn$ is varied solely in the plane of the SOC layer. 

\textit{The lattice Bogoliubov--de Gennes framework}.--- 
We consider a 3D cubic S/SOC lattice structure of size $N_x \times N_y \times N_z$ with interface normal along the $x$ axis. We assume periodic boundary conditions along the $y$ and $z$ axes. The inversion symmetry-breaking in the nonsuperconducting layer is accounted for by the existence of a Rashba SOC term in the Hamiltonian, with a constant magnitude $\lambda$. In addition, we include a perpendicular Rashba contribution with $\vecn_{\text{int}}=\boldsymbol{x}$ and magnitude $\lambda_{\text{int}}$ at the atomic layer closest to the interface. Our Hamiltonian thus accounts for both a Rashba-like SOC field in the bulk of the nonsuperconducting material, and interfacial Rashba SOC.
We use the ballistic-limit tight-binding Bogoliubov--de Gennes framework, following a similar approach to that in Refs.~\cite{terrade_thesis_15, linder_prb_17, johnsen_prb_19}. Our Hamiltonian is given by
\begin{equation}
    \begin{split}
        \label{Hamiltonian}
        H = &-t\sum_{\left<\boldsymbol{i},\boldsymbol{j}\right>,\sigma}c_{\boldsymbol{i},\sigma}^\dagger c_{\boldsymbol{j},\sigma}
        -\sum_{\boldsymbol{i},\sigma} \mu_{\boldsymbol{i}} c_{\boldsymbol{i},\sigma}^\dagger c_{\boldsymbol{i},\sigma}\\
        &-\sum_{\boldsymbol{i}} U_{\boldsymbol{i}}n_{\boldsymbol{i},\uparrow}n_{\boldsymbol{i},\downarrow} 
        -\frac{i}{2}\sum_{\left<\boldsymbol{i},\boldsymbol{j}\right>,\alpha,\beta}c_{\boldsymbol{i},\alpha}^{\dagger}(\lambda\vecn+\lambda_{\text{int}}\vecn_{\text{int}}) \\
        &\cdot\Big\{\boldsymbol{\sigma}\times\Big[\frac{1}{2}(1+\zeta)(\boldsymbol{d}_{\boldsymbol{i},\boldsymbol{j}})_{x}
        +(\boldsymbol{d}_{\boldsymbol{i},\boldsymbol{j}})_{||}\Big]\Big\}_{\alpha,\beta}c_{\boldsymbol{j},\beta}.
    \end{split}
\end{equation}
Above, $t$ is the hopping integral, $\mu_{\boldsymbol{i}}$ is the chemical potential at lattice site $\boldsymbol{i}$, $U_{\boldsymbol{i}}>0$ is the attractive on-site interaction giving rise to superconductivity, $\boldsymbol{\sigma}$ is the vector of Pauli matrices, $\boldsymbol{d}_{\boldsymbol{i},\boldsymbol{j}}$ is the vector from site $\boldsymbol{i}$ to site $\boldsymbol{j}$, and $(\boldsymbol{d}_{\boldsymbol{i},\boldsymbol{j}})_{x}$ and $(\boldsymbol{d}_{\boldsymbol{i},\boldsymbol{j}})_{||}$ are its projections onto the $x$ axis and $yz$ plane, respectively. If site $\boldsymbol{i}$ and $\boldsymbol{j}$ are both inside the SOC layer, $\zeta=1$. Otherwise, $\zeta=0$.
$c_{\boldsymbol{i},\sigma}^\dagger$ and $c_{\boldsymbol{i},\sigma}$ are the second quantization electron creation and annihilation operators at site $\boldsymbol{i}$ with spin $\sigma$, and $n_{\boldsymbol{i},\sigma}\equiv c_{\boldsymbol{i},\sigma}^\dagger c_{\boldsymbol{i},\sigma}$ is the number operator. 
The Rashba term \cite{bychkov_jpc_84} has been symmetrized in order to allow for IP components of $\vecn$ while ensuring a Hermitian Hamiltonian.
The superconducting term is treated by a mean-field approach, assuming $c_{\boldsymbol{i},\uparrow} c_{\boldsymbol{i},\downarrow} = \left< c_{\boldsymbol{i},\uparrow} c_{\boldsymbol{i},\downarrow} \right> +\delta$ and neglecting terms of second order in the fluctuations $\delta$. 
The terms of the Hamiltonian are only nonzero in their respective regions.

\begin{figure}[ht]
    \centering
    \includegraphics[width=\columnwidth]{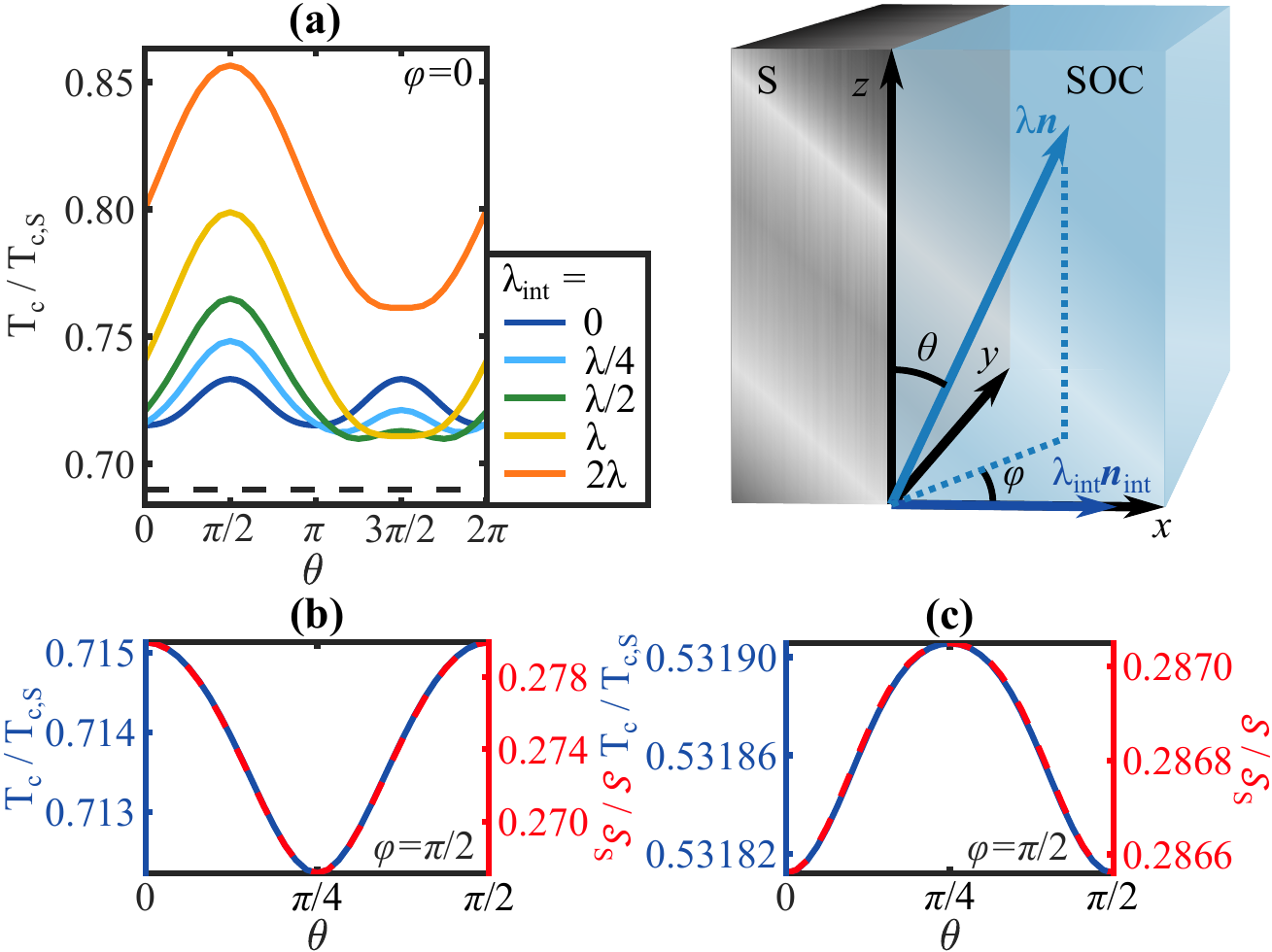}
    \caption{The $T_c$ modulation under rotation of $\vecn$ between IP and OOP orientations (a) is qualitatively different for $\lambda_{\text{int}}<\lambda$ and $\lambda_{\text{int}}>\lambda$. The dashed line marks $T_c$ for $\lambda=\lambda_{\text{int}}=0$. Depending on the material parameters, $T_c$ can have either its IP maxima (b) or minima (c) along the cubic axes. Notice the strong correlation between $T_c$ and the total $s$-wave singlet amplitude at $T=T_c^-$. Above, $T_{c,S}$ and $\mathcal{S}_S$ corresponds to when the superconductor is without proximity to the SOC layer. We have used parameters $N_{x,S}=7$, $N_{x,HM}=3$, $N_y = N_z = 85$, $\mu_S = 1.9$, $\mu_{HM} =1.7$, $U=2.1$, $\lambda=0.8$, $N_T =20$, and $N_{\Delta}=35$ for panels (a) and (b), and $N_{x,S}=5$, $N_{x,HM}=2$, $N_y =N_z =100$, $\mu_S =1.9$, $\mu_{HM}=1.7 $, $U=1.9$, $\lambda=0.2$, $N_T =25$, and $N_{\Delta}=40$ for panel (c), corresponding to coherence lengths $\xi=4$ and $\xi=7$, respectively. In panels (b) and (c), $\lambda_{\text{int}}=0$. }
    \label{fig:Tc}
\end{figure}

We diagonalize the Hamiltonian numerically and compute the physical quantities of interest as outlined in the Supplemental Material.
The superconducting gap $\Delta_{\boldsymbol{i}} \equiv U_{\boldsymbol{i}} \left<c_{\boldsymbol{i},\uparrow}c_{\boldsymbol{i},\downarrow}\right>$ is treated iteratively.
We calculate $T_c$ by a binomial search \cite{ouassou_thesis_15} where we for each of the $N_T$ temperatures considered decide whether the gap increases toward a superconducting state or decreases toward a normal state from an initial guess much smaller than the zero-temperature gap. In this way, we do not calculate the exact value for the gap, and we can thus get high accuracy in $T_c$ for a low number of iterations $N_{\Delta}$.

In order to confirm that the modulation of $T_c$ is caused by conversion of $s$-wave even-frequency singlets into other singlet and triplet correlations, we consider the even-frequency $s$-wave singlet amplitude $S_{s,\boldsymbol{i}}\equiv\left<c_{\boldsymbol{i},\uparrow}c_{\boldsymbol{i},\downarrow}\right>-\left<c_{\boldsymbol{i},\downarrow}c_{\boldsymbol{i},\uparrow}\right>$. As a measure of the total $s$-wave singlet amplitude of the superconductor, we introduce the quantity $\mathcal{S} \equiv\frac{1}{N_{x,S}}\sum_{i}|S_{s,i}|$, where the sum is taken over the superconducting region only. 
We also define the opposite- and equal-spin odd-frequency $s$-wave triplet amplitudes $S_{0,\boldsymbol{i}}(\tau)\equiv\left<c_{\boldsymbol{i},\uparrow}(\tau)c_{\boldsymbol{i},\downarrow}(0)\right>+\left<c_{\boldsymbol{i},\downarrow}(\tau)c_{\boldsymbol{i},\uparrow}(0)\right>$, and $S_{\sigma,\boldsymbol{i}}(\tau)\equiv\left<c_{\boldsymbol{i},\sigma}(\tau)c_{\boldsymbol{i},\sigma}(0)\right>$ \cite{linder_prb_17},
where the time-dependent electron annihilation operator is given by $c_{\boldsymbol{i},\sigma}(\tau)\equiv e^{iH\tau}c_{\boldsymbol{i},\sigma}e^{-iH\tau}$ \cite{fetter_book_03}. 
The $s$-wave triplet amplitude is of particular interest as it is the only triplet amplitude robust to impurity scattering. Other superconducting correlations, such as $p$-wave and $d$-wave correlations, also appear due to the presence of SOC, as will be discussed later in this work.

\textit{The superconducting critical temperature}.--- By following the above approach, we plot the critical temperature and the total $s$-wave singlet amplitude in Fig.~\ref{fig:Tc}. To ensure that the effect is robust, we use two different parameter sets. The parameters are given in the figure caption.
All length scales are scaled by the lattice constant $a$, the SOC magnitudes are scaled by $ta$, and the remaining energy scales are scaled by $t$. For $t\sim1$ eV and $a\sim5$ \AA, the order of magnitude of $\lambda$ is $10^{-10}$ eVm, which corresponds well to Rashba parameters found in several materials \cite{manchon_nm_15}. In order to make the system computationally manageable, the lattice size and coherence length $\xi\propto\Delta^{-1}$ must be scaled down, leading to an overestimation of $\Delta$ and thus $T_c$. The results in Fig.~\ref{fig:Tc} must therefore be seen mainly as qualitative. 

For both sets of parameters, we see a qualitatively similar behavior for rotations of $\vecn$ in the $xz$ plane, (see Fig.~\ref{fig:Tc}(a) for the first set of parameters).
When $\lambda_{\text{int}}=0$, we find a suppression of $T_c$ for an IP $\vecn$ compared to an OOP $\vecn$.
When $0<\lambda_{\text{int}}<\lambda$, there are still maxima at the OOP directions $\vecn||\vecn_{\text{int}}$ and $(-\vecn)||\vecn_{\text{int}}$, but when increasing $\lambda_{\text{int}}$ the magnitude of the former increases while the magnitude of the latter decreases. As long as $\vecn$ is parallel to $\vecn_{\text{int}}$ in the OOP configuration, the $T_c$ modulation from IP to OOP is thus enhanced by the additional interface contribution. 
For $\lambda_{\text{int}}>\lambda$, $T_c$ is maximal for $\vecn||\vecn_{\text{int}}$ and minimal for $(-\vecn)||\vecn_{\text{int}}$. The change in $T_c$ from the parallel to the anti parallel configuration increases with an increasing $\lambda_{\text{int}}$. The results presented here only depend on the relative orientations of $\vecn$ and $\vecn_{\text{int}}$, and are independent of whether $\vecn_{\text{int}}$ is directed out of or into the nonsuperconducting material. 
Notice that in all cases, nonzero SOC increases $T_c$ compared to when $\lambda=\lambda_{\text{int}}=0$. This is explained by a decreased leakage of conventional singlets into the nonsuperconducting region. 

From panels (b) and (c), we see that there is also an IP variation in $T_c$, that may give the strongest in-plane suppression either when $\vecn$ is oriented at a $\pi/4$ angle with respect to the cubic axes, or when $\vecn$ is oriented along the cubic axes. As we find a similar variation in the normal-state free energy, which only depends on the eigenenergy spectrum of the system, this varying modulation of the IP component of $T_c$ is likely to be caused by band-structure effects due to the crystal structure of the cubic lattice. In order to demonstrate the IP modulation, the interfacial SOC should preferably be as small as possible.

To demonstrate that the $T_c$ modulation can be attributed to the variation of the $s$-wave singlet amplitude in the superconducting region, we plot the total $s$-wave singlet amplitude as a function of the IP angle of $\vecn$ (panels (b) and (c)). As expected, it is of a similar form as the variation in $T_c$. The slight deviation between $T_c$ and $\mathcal{S}$ is caused by $\mathcal{S}$ being calculated at a temperature $T_c^-$ slightly below $T_c$. We have verified that the variation in $\mathcal{S}$ and $T_c$ is similar also for panel (a).

The variation in the $s$-wave singlet amplitude inside the superconducting region is caused by a reduced leakage of $s$-wave singlets out of the superconducting region, and conversion of $s$-wave singlets into other singlet and triplet correlations. 
When $\lambda_{\text{int}}$ is nonzero, the length of $\lambda\vecn+\lambda_{\text{int}}\vecn_{\text{int}}$ changes under rotations of $\vecn$, leading to an effective change in the magnitude of the SOC. Increased SOC causes an increase in the Fermi vector mismatch \cite{tanaka_prb_90}, due to a change in the Fermi surface in the nonsuperconducting material. Since the overlap between the Fermi surfaces of the two materials decreases, there is an increase in the normal reflection at the interface, as our analytical results verify.
For large $\lambda_{\text{int}}$, the $T_c$ modulation is dominated by variation in the Fermi vector mismatch. 
If we further investigate the triplet amplitudes present for different orientations of $\vecn$, we find that the $s$-wave odd-frequency triplet amplitude is absent for $\vecn=\boldsymbol{x}$, \textit{i.e.} when $\vecn$ has no IP component. For all other orientations of $\vecn$, the $s$-wave odd-frequency anomalous triplet amplitude is nonzero. This suggests that the OOP to IP change in $T_c$ is at least partly caused by the increase in the $s$-wave triplet amplitude from zero when $\vecn$ points OOP to an increasing finite value as the IP component of $\vecn$ increases.
When $\lambda_{\text{int}}$ is small, so that the length of $\lambda\vecn+\lambda_{\text{int}}\vecn_{\text{int}}$ is approximately constant, we may therefore expect an IP suppression of $T_c$ not only in the ballistic-limit materials covered by our theoretical framework, but also in diffusive materials. 
Below, we perform analytical calculations which prove that odd-frequency pairing is absent when $\vecn$ points OOP.

\textit{The continuum Bogoliubov--de Gennes framework}.--- In order to explain the absence of $s$-wave odd-frequency triplets when $\vecn$ is OOP, we consider two 2D continuum systems that can be treated analytically within the Bogoliubov--de Gennes framework \cite{mcmillan_pr_67,ishii_ptp_70,furusaki_ssc_91,kashiwaya_rpp_00,koperdraad_jp_01,lu_ptrsa_16,cayao_prb_17,cayao_prb_18}: a SOC/S bilayer with an OOP $\vecn = \boldsymbol{x}$, and a F/S bilayer with magnetization $\vecm \parallel \boldsymbol{z}$. We use conventions similar to those in Refs. \cite{cayao_prb_17,cayao_prb_18}. Our systems are located in the $xy$ plane, with interface normal along $\vecx$ and the interface at $x=0$. 

We find the scattering wave functions $\Psi_n (x_1 )$, and $\Tilde{\Psi}_m (x_2 )$ that we will use to construct the Green's functions in the system from the time-independent Schr\"{o}dinger equations \cite{blonder_prb_82,cayao_prb_17,cayao_prb_18} 
\begin{equation}
    \begin{split}
        H(p_y)\Psi_n (x_1 )&=(\omega+i\delta)\Psi_n (x_1 ),\\
        H^* (p_y)\Tilde{\Psi}_m (x_2 )&=(\omega+i\delta)\Tilde{\Psi}_m (x_2),
    \end{split}
\end{equation}
respectively, where
\begin{equation}
    \begin{split}
    &H(p_y)=(-\partial_{x} ^2 /\eta+p_y^2 /\eta-\mu)\hat{\tau}_3 \hat{\sigma}_0 \\
    &+\Delta i\hat{\tau}^+ \hat{\sigma}_y -\Delta^{*} i\hat{\tau}^- \hat{\sigma}_y +h_x \hat{\tau}_3 \hat{\sigma}_x +h_y \hat{\tau}_0 \hat{\sigma}_y + h_z \hat{\tau}_3 \hat{\sigma}_z \\
    &-\lambda ( n_x p_y +n_y i\partial_x )\hat{\tau}_0 \hat{\sigma}_z+i\lambda n_z \partial_x \hat{\tau}_3 \hat{\sigma}_y +\lambda n_z p_y \hat{\tau}_0 \hat{\sigma}_x.\\
    \end{split}
    \label{Hamiltonian_continuum}
\end{equation}
Above, $\delta>0$ is real and infinitesimal, $\eta\equiv2m/\hbar^2$, $p_y$ is the momentum in the $y$ direction, and $\boldsymbol{h}=(h_x ,h_y ,h_z )$ is the magnetic exchange field. The terms are only nonzero in their respective regions.
The four components of the scattering wave functions correspond to spin-up and spin-down electrons, and spin-up and spin-down holes, respectively. The spins are defined with respect to the $z$ axis.
The indices $n$ and $m$ refer to the eight possible wave functions describing scattering of quasiparticles incoming from the left and right. 
In the continuum model, the symmetrization of the Rashba term enters through the boundary conditions of the wave functions at $x=0$ rather than through the Hamiltonian \cite{reeg_prb_15}.
From the scattering wave functions, we construct the retarded Green's function in Nambu~$\otimes$~spin space  for $x_1 > x_2$ and $x_1 < x_2$, and apply boundary conditions at $x_1 =x_2$.

The even-(odd-)frequency singlet and triplet retarded anomalous Green's functions can be written in terms of the center of mass coordinate $X\equiv(x_1 +x_2 )/2$ and the relative coordinate $x\equiv x_1 -x_2$ as \cite{cayao_prb_17,cayao_prb_18}
\begin{equation}
\begin{split}
    F_0^{r,E(O)}(X,x ,p_y ;\omega)&=[F_{0}^{r} (X,x ,p_y ;\omega)\\
    &\varpm F_{0}^{r} (X,-x ,-p_y  ;\omega)]/2,\\
    F_{i}^{r,E(O)} (X,x ,p_y ;\omega)&=[F_{i}^{r} (X,x ,p_y ;\omega)\\
    &\varmp F_{i}^{r} (X,-x ,-p_y ;\omega)]/2,\\
\end{split}
\label{fEO}
\end{equation}
where $i=\{1,2,3\}$, and
\begin{equation}
    \begin{split}
    F_0^{r} (X,x ,p_y ;\omega)&=[F_{\uparrow\downarrow}^{r} (X,x ,p_y ;\omega)
    -F_{\downarrow\uparrow}^{r} (X,x ,p_y ;\omega)]/2,\\
    F_1^{r} (X,x ,p_y  ;\omega)&=\phantom{[}F_{\uparrow\uparrow}^{r} (X,x ,p_y ;\omega),\\
    F_2^{r} (X,x ,p_y ;\omega)&=\phantom{[}F_{\downarrow\downarrow}^{r} (X,x ,p_y  ;\omega),\\
    F_3^{r} (X,x ,p_y ;\omega)&=[F_{\uparrow\downarrow}^{r} (X,x ,p_y  ;\omega)
    +F_{\downarrow\uparrow}^{r} (X,x ,p_y ;\omega)]/2
    \end{split}
\end{equation}
represents the singlet amplitude, the equal-spin triplet amplitudes ($i=1,2$), and the opposite-spin triplet amplitude ($i=3$), respectively. The retarded anomalous Green's functions $F_{\sigma\sigma'}^r (X,x ,p_y ;\omega)$ are anomalous elements of the retarded Green's function in Nambu~$\otimes$~spin space.
Odd (even) frequency refers to the symmetry of the Green's function under inversion of relative time, corresponding to $\omega$ changing (not changing) sign.

The analytical expressions obtained for the even- and odd-frequency singlet and triplet retarded anomalous Green's functions are given in the Supplemental Material. Their spatial symmetries are determined by their parities under inversion of $x$ and $p_y$. Although the $s$-wave and $d_{x^2 -y^2}$-wave triplets have the same parities along the $x$ and $y$ axis, the presence of the $s$-wave triplet is proven by a nonzero result when integrating over all spatial coordinates.

\textit{Singlet and triplet amplitudes}.--- 
For the 2D SOC/S structure with $\vecn=\boldsymbol{x}$, we find that $s$- and $p_x$-wave singlets, and $p_y$- and $d_{xy}$-wave opposite-spin triplets are present.
At the first glance, it might seem strange that the odd-frequency $s$-wave triplet amplitude is zero, when it is nonzero for a 2D F/S structure with magnetization along the $z$ axis. Although the Hamiltonians of these systems are of a similar form, they allow for the existence of different triplet amplitudes. The crucial difference leading to a generation of $p_y$- and $d_{xy}$-wave triplets in the SOC/S system rather than $s$- and $p_x$-wave triplets as in the F/S system, is the momentum dependence of the Rashba term.

We have also investigated a 2D SOC/S structure for an IP orientation $\vecn = \boldsymbol{z}$ numerically and find additional equal-spin triplets with an odd-frequency symmetry. For a 3D SOC/S system with $\vecn$ OOP, the Rashba term depends on the momentum both along the $y$ and $z$ axes. Similarly as in 2D, we expect this to allow for triplets that are odd under inversion of $p_y$ and $p_z$. This is ultimately the reason for the absence of $s$-wave triplets.

\begin{figure}
    \centering
    \includegraphics[width=\columnwidth]{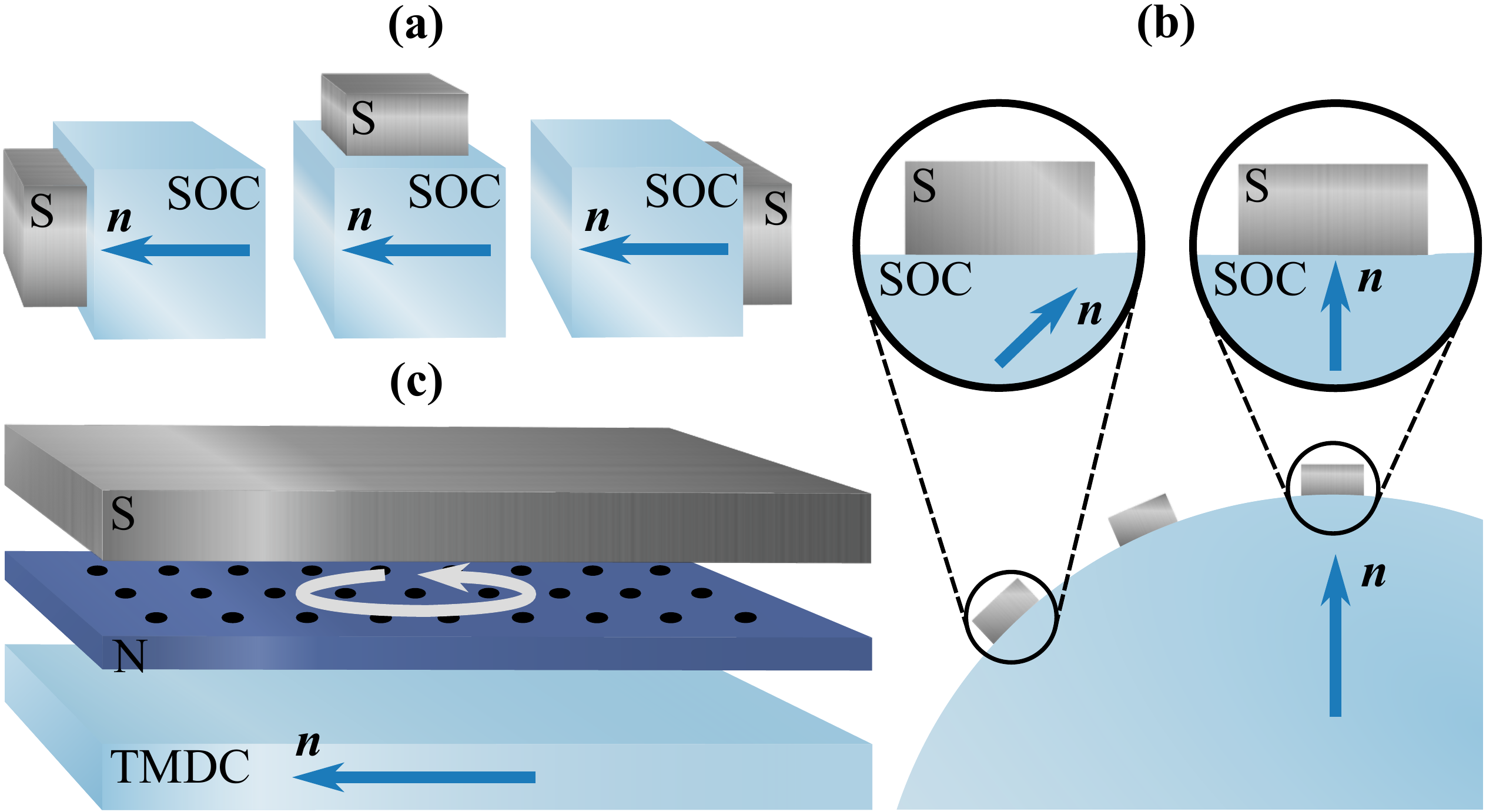}
    \caption{For the experimental observation of the IP to OOP $T_c$ modulation, we suggest growing the superconductor on (a) different surfaces of a non centrosymmetric material or (b) on a curved non centrosymmetric material. For observing IP variations, we suggest (c) growing a normal metal with a cubic lattice structure at different angles compared to a TMDC with IP inversion symmetry-breaking, and then growing the superconductor on top. The N/TMDC bilayer effectively enables a rotation of $\vecn$ compared to the lattice.}
    \label{fig:exp}
\end{figure}

\textit{Experimental realization}.--- We finally comment on the possibilities of an experimental realization of the predicted $T_c$ variation upon redirecting $\vecn$. We suggest cleaving a non centrosymmetric metal, such as BiPd \cite{zhuravlev_spj_57, bhatt_jlcm_79, sun_nc_15}, in different directions and growing a superconductor (with a higher $T_c$) on the surface, see Fig.~\ref{fig:exp}(a). This requires a material that can be cleaved along at least two axes.
Alternatively, one could deposit superconductors on the surface of a curved non centrosymmetric material with a long edge (several mm), see Fig.~\ref{fig:exp}(b) \cite{ortega_prb_11}. In both scenarios, different samples would have their inversion symmetry-breaking axis in different directions, corresponding to a systematic rotation of $\vecn$ from IP to OOP. 
We underline that although $\vecn$ rotates along with the lattice in the nonsuperconducting region, the difference in $T_c$ as $\vecn$ changes from IP to OOP is robust. The reason is that the corresponding change in the proximity effect exists even in our continuum model without the underlying lattice.

In order to observe IP variations, we suggest growing a normal metal (N) with a cubic lattice structure at different angles compared to a transition metal dichalcogenide (TMDC) with IP inversion symmetry-breaking \cite{absor_jap_17}, see Fig.~\ref{fig:exp}(c). This corresponds to an effective IP rotation of $\vecn$ compared to the lattice. The superconductor is grown on top of the normal metal, which should be a light element with as little interfacial SOC as possible. 
The ideal scenario, albeit challenging, would be to induce an \textit{in situ} rotation of $\vecn$ in the nonsuperconducting region via electric gating in different directions, that induces an inversion-symmetry-breaking field. However, since $\vecn$ is rotated inside the non centrosymmetric material, $\lambda$ may in principle vary. This is not the case for our previous suggestions, since we do not rotate $\vecn$ inside the non centrosymmetric material, but instead change the position of the superconductor.

Concluding, we have shown that the superconducting transition temperature $T_c$ can be altered by rotating the inversion symmetry-breaking axis $\vecn$ in a proximate material, providing a conceptually different way of controlling $T_c$ compared to previous studies. Moreover, we have shown that when in addition an interfacial spin-orbit coupling perpendicular to the interface is present and substantial, the behavior of $T_c$ as $\vecn$ is varied can change qualitatively.

\vspace{1em}
\noindent
The authors would like to thank J. A. Ouassou, and J. W. Wells for helpful discussions. This work was supported by the Research Council of Norway through its Centres of Excellence funding scheme Grant No. 262633 QuSpin.
\vspace{2em}

\begin{center}
    \noindent
{\large \textbf{Supplemental Material}}
\end{center}

\section{The lattice Bogoliubov--de Gennes framework}

If the inversion symmetry-breaking axis directed along $\vecn$ has an in-plane component, a Rashba Hamiltonian of the form 
\begin{equation}
    -\frac{i}{2}\sum_{\left< \boldsymbol{i},\boldsymbol{j}\right> ,\alpha,\beta} \lambda_{\boldsymbol{i}} c_{\boldsymbol{i},\alpha}^{\dagger} \vecn \cdot (\boldsymbol{\sigma}\times\boldsymbol{d}_{\boldsymbol{i},\boldsymbol{j}})_{\alpha,\beta} c_{\boldsymbol{j},\beta}
\end{equation}
is in general non-Hermitian. This term is the second quantized form of $\hat{h}=(\vecn \times\boldsymbol{\sigma})\cdot\lambda(x)\hat{\boldsymbol{p}}$, where $\boldsymbol{\sigma}$ is the vector of Pauli matrices, $\lambda(x)$ is the $x$ dependent Rashba spin-orbit coupling strength, and $\hat{\boldsymbol{p}}=(\hat{p}_x , \hat{p}_y ,\hat{p}_z )= -i\hbar\nabla$ is the momentum operator \cite{bychkov_jpc_84}. Above, $\boldsymbol{d}_{\boldsymbol{i},\boldsymbol{j}}$ is the vector from lattice site $\boldsymbol{i}$ to site $\boldsymbol{j}$.
More generally, the symmetrized version of the first quantized Rashba spin-orbit coupling operator is
\begin{equation}
    \hat{h}=\frac{1}{2}(\vecn \times\boldsymbol{\sigma})\cdot\{\lambda(x),\hat{\boldsymbol{p}}\}.
    \label{h_symmetrised}
\end{equation}
We write this on a second quantized form as $H_{\lambda}=\sum_{\boldsymbol{i},\boldsymbol{j},\alpha,\beta}\big<\boldsymbol{i},\alpha\big|\hat{h}\big|\boldsymbol{j},\beta\big>c^{\dagger}_{\boldsymbol{i},\alpha} c_{\boldsymbol{j},\beta}$. The spatial part of the overlap integral can be written
\begin{equation}
\begin{split}
    \big<\boldsymbol{i}\big|\hat{h}\big|\boldsymbol{j}\big>=&\hspace{1mm}\frac{1}{2}\hspace{0.5mm}(\vecn \times\boldsymbol{\sigma})\cdot\boldsymbol{x}\big[\big<\boldsymbol{i}\big|\lambda(x)\hat{p}_x \big|\boldsymbol{j}\big> + \big<\boldsymbol{j}\big|\lambda(x)\hat{p}_x \big|\boldsymbol{i}\big>^{*} \big]\\
    &+(\vecn \times\boldsymbol{\sigma})\cdot\boldsymbol{y}\big<\boldsymbol{i}\big|\lambda(x)\hat{p}_y\big|\boldsymbol{j}\big>\\
    &+(\vecn \times\boldsymbol{\sigma})\cdot\boldsymbol{z}\big<\boldsymbol{i}\big|\lambda(x)\hat{p}_z\big|\boldsymbol{j}\big>,
\end{split}
\label{overlap_reformulated}
\end{equation}
where 
\begin{equation}
    \big<\boldsymbol{i}\big|\lambda(x)\hat{p}_m \big|\boldsymbol{j}\big>=\int_{-\infty}^{\infty}dm\hspace{1mm}\phi_{\boldsymbol{i}}^* (\boldsymbol{r})\lambda(x) \hat{p}_m \phi_{\boldsymbol{j}}(\boldsymbol{r})
\end{equation}
for $m=x,y,z$. Here, $\phi_{\boldsymbol{j}}(\boldsymbol{r})\equiv\phi(\boldsymbol{r}-\boldsymbol{R}_{\boldsymbol{j}})$, where $\boldsymbol{R}_{\boldsymbol{j}}$ describes the position of lattice site $\boldsymbol{j}$. We assume each $\phi_{\boldsymbol{j}}$ to be highly localized. Then $\lambda(x)$ can be approximated to be constant inside each Wigner-Seitz cell, the derivative can be discretized as
\begin{equation}
    \partial_m \phi_{\boldsymbol{j}}(\boldsymbol{r})=\frac{1}{2}[\phi_{\boldsymbol{j-\hat{m}}}(\boldsymbol{r})-\phi_{\boldsymbol{j+\hat{m}}}(\boldsymbol{r})],
\end{equation}
and
\begin{equation}
    \int d\boldsymbol{r} \phi_{\boldsymbol{i}}^{*}(\boldsymbol{r})\phi_{\boldsymbol{j}}(\boldsymbol{r})=\delta_{\boldsymbol{i},\boldsymbol{j}}.
\end{equation}
We also assume that $\lambda(x)=\lambda$ is constant and nonzero inside the material with spin-orbit coupling, and that $\lambda(x)$ acts as a step function at the interface. 
It follows that the symmetrized spin-orbit coupling contribution to the Hamiltonian is
\begin{equation}
    \begin{split}
        H_{\lambda}=&-\frac{i}{2}\sum_{\left<\boldsymbol{i},\boldsymbol{j}\right>,\alpha,\beta}\lambda c_{\boldsymbol{i},\alpha}^{\dagger}\vecn \\
        &\cdot\Big\{\boldsymbol{\sigma}\times\Big[\frac{1}{2}(1+\zeta)(\boldsymbol{d}_{\boldsymbol{i},\boldsymbol{j}})_{x}
        +(\boldsymbol{d}_{\boldsymbol{i},\boldsymbol{j}})_{||}\Big]\Big\}_{\alpha,\beta}c_{\boldsymbol{j},\beta}.
    \end{split}
    \label{SOC_H1}
\end{equation}
Above, $\boldsymbol{d}_{\boldsymbol{i},\boldsymbol{j}}$ is decomposed into a part $(\boldsymbol{d}_{\boldsymbol{i},\boldsymbol{j}})_{x}$ perpendicular to the interface, and a part $(\boldsymbol{d}_{\boldsymbol{i},\boldsymbol{j}})_{||}$ parallel to the interface. If site $\boldsymbol{i}$ and $\boldsymbol{j}$ are both inside the material with spin-orbit coupling, $\zeta=1$, while if site $\boldsymbol{i}$ and site $\boldsymbol{j}$ are on opposite sides of the interface, $\zeta=0$.

Using the symmetrized Rashba contribution, our Hamiltonian is given by Eq.~(1) in the letter. In the following, we use a similar approach to that in Refs.~\cite{terrade_thesis_15, linder_prb_17, johnsen_prb_19}. For brevity of notation, we introduce $i\equiv i_x$, $\boldsymbol{i}_{||}\equiv(i_y ,i_z )$, and $\boldsymbol{k}\equiv(k_y ,k_z )$.
We assume periodic boundary conditions in the $y$ and $z$ directions, and introduce the Fourier transform along the $y$ and $z$ axes,
\begin{equation}
    c_{\boldsymbol{i},\sigma}=\frac{1}{\sqrt{N_y N_z}}\sum_{\boldsymbol{k}} c_{i,\boldsymbol{k},\sigma} e^{i(\boldsymbol{k}\cdot \boldsymbol{i}_{||})},
\end{equation}
where the sum is taken over the allowed $\boldsymbol{k}$ inside the first Brillouin zone. In the following, we also use the relation 
\begin{equation}
    \frac{1}{N_y N_z}\sum_{\boldsymbol{i}_{||}} e^{i(\boldsymbol{k}-\boldsymbol{k} ')\cdot\boldsymbol{i}_{||}} =\delta_{\boldsymbol{k} , \boldsymbol{k} '}.
    \label{trick}
\end{equation}
By choosing the basis 
\begin{equation}
    B_{i , \boldsymbol{k} }^{\dagger}\equiv[c_{i , \boldsymbol{k}   ,\uparrow}^{\dagger} \hspace{3mm} c_{i , \boldsymbol{k}   ,\downarrow}^{\dagger} \hspace{3mm} c_{i , - \boldsymbol{k} ,\uparrow} \hspace{3mm} c_{i , -\boldsymbol{k} ,\downarrow}],
\end{equation}
and applying the Fourier transform as well as Eq.~\eqref{trick}, we rewrite the Hamiltonian as 
\begin{equation}
    H=H_0+\frac{1}{2}\sum_{i , j , \boldsymbol{k}  } B_{i , \boldsymbol{k}  }^\dagger H_{i , j , \boldsymbol{k}  }B_{i , \boldsymbol{k}  }.
\end{equation}
The constant term $H_0$ is of no importance for our further calculations. Above,
\begin{align}
\label{Hamiltonian2}
    H_{i , j , \boldsymbol{k}  }&= \epsilon_{i ,j ,\boldsymbol{k}  }\hat{\tau}_3 \hat{\sigma}_0 
    \notag\\
		&+(\Delta_{i}\hat{\tau}^+  -\Delta_{i}^* \hat{\tau}^- )i\hat{\sigma}_y \delta_{i ,j}\notag\\
    &-\{(\lambda  n_x +\lambda_{\text{int}} |\vecn_{\text{int}}|) [\sin(k_y ) \hat{\tau}_0 \hat{\sigma}_z  -\sin(k_z ) \hat{\tau}_3 \hat{\sigma}_y ]\notag\\
    &-\lambda[n_z \sin(k_y) -n_y \sin(k_z ) ]\hat{\tau}_0 \hat{\sigma}_x \}\delta_{i ,j}\notag\\
    &+i\lambda(1+\zeta)(n_y \hat{\tau}_0  \hat{\sigma}_z 
    - n_z \hat{\tau}_3 \hat{\sigma}_y   )\notag\\
    &\cdot(\delta_{i ,j +1}-\delta_{i ,j -1})/4,
\end{align}
where $\hat{\tau}^\pm \equiv(\hat{\tau}_1 \pm i\hat{\tau}_2 )/2$, $\hat{\tau}_i \hat{\sigma}_j \equiv \tau_i \otimes\sigma_j$ is the Kronecker product of the Pauli matrices spanning Nambu and spin space,
\begin{equation}
\begin{split}
    \epsilon_{i , j , \boldsymbol{k}  } &\equiv \{-2t[\cos(k_y)+\cos(k_z)]-\mu_{i}\}\delta_{i , j}\\
    &-t(\delta_{i , j +1}+\delta_{i , j -1} ),
\end{split}
\end{equation}
and $\Delta_{i}$ is the superconducting gap at site $i$.
By rewriting the Hamiltonian as 
\begin{equation}
    H=H_0 + \frac{1}{2}\sum_{\boldsymbol{k}  }W_{\boldsymbol{k}  }^\dagger H_{\boldsymbol{k}  } W_{\boldsymbol{k}  }
\end{equation}
in terms of the basis
\begin{equation}
    W_{\boldsymbol{k}  }^\dagger \equiv [B_{1,\boldsymbol{k}  }^\dagger,...,B_{i ,\boldsymbol{k}  }^\dagger ,...,B_{N_x ,\boldsymbol{k}  }^\dagger ],
\end{equation}
the Hamiltonian can be diagonalized numerically as
\begin{equation}
    H=H_0+\frac{1}{2}\sum_{n, \boldsymbol{k}  }E_{n, \boldsymbol{k}  }\gamma_{n, \boldsymbol{k}  }^\dagger \gamma_{n, \boldsymbol{k}  }.
\end{equation}
This yields eigenenergies $E_{n, \boldsymbol{k}  }$, and eigenvectors $\Phi_{n,\boldsymbol{k}  }$ given by 
\begin{equation}
    \begin{split}
        \Phi_{n,\boldsymbol{k}  }^{\dagger}&\equiv[ \phi_{1,n,\boldsymbol{k}  }^{\dagger} \hspace{3mm} \cdots \hspace{3mm}\phi_{N_x ,n,\boldsymbol{k}  }^{\dagger}],\\
        \phi_{i ,n,\boldsymbol{k}  }^{\dagger}&\equiv[u_{i ,n,\boldsymbol{k}  }^{*}\hspace{1mm} v_{i ,n,\boldsymbol{k}  }^{*}\hspace{1mm} w_{i ,n,\boldsymbol{k}  }^{*}\hspace{1mm} x_{i ,n,\boldsymbol{k}  }^{*}].\\
    \end{split}
\end{equation}
The new quasiparticle operators introduced above are related to the old operators by 
\begin{equation}
\begin{split}
    c_{i ,\boldsymbol{k}   ,\uparrow}&=\sum_n u_{i ,n, \boldsymbol{k}  }\gamma_{n,\boldsymbol{k}  },\\
    c_{i ,\boldsymbol{k}   ,\downarrow}&=\sum_n v_{i ,n, \boldsymbol{k}  }\gamma_{n,\boldsymbol{k}  },\\
    c^{\dagger}_{i ,-\boldsymbol{k} ,\uparrow}&=\sum_n w_{i ,n, \boldsymbol{k}  }\gamma_{n,\boldsymbol{k}  },\\
    c^{\dagger}_{i ,-\boldsymbol{k} ,\downarrow}&=\sum_n x_{i ,n, \boldsymbol{k}  }\gamma_{n,\boldsymbol{k}  }.\\
\end{split}
\end{equation}
The eigenenergies and eigenvectors are used for computing the singlet and triplet amplitudes and the superconducting critical temperature. In finding the eigenenergies and eigenvectors, the superconducting gap must be calculated iteratively. The superconducting gap is defined by $\Delta_{\boldsymbol{i}} \equiv U_{\boldsymbol{i}} \left<c_{\boldsymbol{i},\uparrow}c_{\boldsymbol{i},\downarrow}\right>$. By Fourier transforming along the $y$ and $z$ axes, rewriting to the new quasi-particle operators, and using that $\langle \gamma_{n,\boldsymbol{k}  }^\dagger \gamma_{m,\boldsymbol{k}  }\rangle =f\big(E_{n,\boldsymbol{k}  }/2\big)\delta_{n,m}$, we find that the gap is given by
\begin{equation}
    \Delta_{i}=-\frac{U_{i}}{N_y N_z}\sum_{n,\boldsymbol{k}  }v_{i ,n,\boldsymbol{k}  }w_{i ,n,\boldsymbol{k}  }^* \left[1-f\left(E_{n,\boldsymbol{k}  }/2\right)\right],
\end{equation}
where $f\big(E_{n,\boldsymbol{k}  }/2\big)$ is the Fermi-Dirac distribution. 

We define the even-frequency $s$-wave singlet amplitude as $S_{s,\boldsymbol{i}}\equiv\left<c_{\boldsymbol{i},\uparrow}c_{\boldsymbol{i},\downarrow}\right>-\left<c_{\boldsymbol{i},\downarrow}c_{\boldsymbol{i},\uparrow}\right>$. The even-frequency $s$-wave singlet amplitude inside the superconducting region is related to the superconducting gap by $S_{s,i}=2\Delta_{i} /U_{i}$.
By the same method as used for finding the expression for the superconducting gap, we find that the odd-frequency $s$-wave triplet amplitudes are given by
\begin{equation}
    \begin{split}
        S_{0,i}(\tau)=&\frac{1}{N_y N_z}\sum_{n,\boldsymbol{k}}[u_{i ,n,\boldsymbol{k}}x_{i ,n,\boldsymbol{k}}^{*}
        +v_{i ,n,\boldsymbol{k}}w_{i ,n,\boldsymbol{k}}^{*}]\\
        &\cdot e^{-iE_{n,\boldsymbol{k}}\tau/2} [1-f(E_{n,\boldsymbol{k}}/2)],\\
        S_{\uparrow,i}(\tau)=&\frac{1}{N_y N_z}\sum_{n,\boldsymbol{k}}u_{i ,n,\boldsymbol{k}}w_{i ,n,\boldsymbol{k}}^{*}\\
        &\cdot e^{-iE_{n,\boldsymbol{k}}\tau/2}[1-f(E_{n,\boldsymbol{k}}/2)],\\
        S_{\downarrow,i}(\tau)=&\frac{1}{N_y N_z}\sum_{n,\boldsymbol{k}}v_{i ,n,\boldsymbol{k}}x_{i ,n,\boldsymbol{k}}^{*}\\
        &\cdot e^{-iE_{n,\boldsymbol{k}}\tau/2}[1-f(E_{n,\boldsymbol{k}}/2)].
    \end{split}
\end{equation}

Our binomial search algorithm \cite{ouassou_thesis_15} for the superconducting critical temperature, is as follows. We divide our temperature interval $N_T$ times. For each of the $N_T$ iterations, we recalculate the gap $N_{\Delta}$ times from an initial guess with a magnitude $\Delta_0 /1000$, where $\Delta_0$ is the zero-temperature superconducting gap. If the gap has increased towards a superconducting solution after $N_{\Delta}$ iterations, we conclude that the current temperature is below $T_c$. If the gap has decreased towards a normal-state solution, we conclude that the current temperature is above $T_c$. We measure the magnitude of the gap in the middle of the superconducting region. The advantage of this algorithm, is that we are not dependent upon recalculating the gap until it converges. The parameter $N_{\Delta}$ must only be large enough that the increase or decrease in $\Delta_{i}$ at site $i=N_{x,S} /2$ reflects the overall behavior of the gap under recalculation. When we choose an initial guess so that the gap as a function of lattice site has a similar shape as for the gap very close to $T_c$, it more likely that the gap increases for all lattice sites, or decreases for all lattice sites, under recalculation. We can then get a high accuracy with a low $N_{\Delta}$.

The superconducting coherence length is given by $\xi\equiv\hbar v_F /\pi\Delta_0$ \cite{bardeen_pr_57}, where $v_F\equiv\frac{1}{\hbar}\frac{dE_{\boldsymbol{k}}}{dk}\big|_{k=k_F}$ is the normal-state Fermi velocity \cite{bardeen_pr_57}, $E_{\boldsymbol{k}}$ is the normal-state eigenenergies when introducing periodic boundary conditions along all three axes, and $k_F$ is the corresponding Fermi momentum averaged over the Fermi surface. We round $\xi$ down to the closest integer number of lattice points. 

\section{The continuum Bogoliubov--de Gennes framework}

The continuum Bogoliubov--de-Gennes framework \cite{mcmillan_pr_67,koperdraad_jp_01,kashiwaya_rpp_00,furusaki_ssc_91,ishii_ptp_70,lu_ptrsa_16,cayao_prb_17,cayao_prb_18} allows us to obtain analytical expressions for the singlet and triplet retarded anomalous Green's functions of the 2D SOC/S system with $\vecn =\boldsymbol{x}$ and the 2D F/S system with $\vecm||\boldsymbol{z}$. We have not given these expressions in the letter, as we are mainly interested in their symmetries under spatial inversion. Here, we provide the analytical expressions for the wave functions and the singlet and triplet retarded anomalous Green's functions for these two systems, as well as the wave functions for the 2D SOC/S system with $\vecn =\boldsymbol{z}$.

\subsection{The scattering wave functions}
\label{s:wavefunctions}

We find expressions for the scattering wave functions $\Psi_n (x_1 )$ and $\Tilde{\Psi}_m (x_2 )$ by using the time-independent Schr\"{o}dinger equations given in Eq.~(2) in the letter. The indices $n$ and $m$ refers to the different possible scattering processes. These contain reflection and transmission coefficients that are determined by the boundary conditions at the interface.
The wave functions $\Psi_n (x_1 )$ satisfies the boundary conditions \cite{reeg_prb_15} 
\begin{equation}
    \begin{split}
        [\Psi_n (x_1 )]_{x_1=0^+}&=[\Psi_n (x_1 )]_{x_1=0^-}\\
        [\hat{v} \Psi_n (x_1)]_{x_1=0^+}&=[\hat{v} \Psi_n (x_1)]_{x_1=0^-}\\
    \end{split}
\end{equation}
where $\hat{v}\equiv\partial H(p_y )/\partial (-i\partial_{x_1} )$ is the velocity operator. The conjugate wave functions $\Tilde{\Psi}_m (x_2)$ satisfies a similar set of boundary conditions with $\hat{v}\equiv\partial H^{*}( p_y ) /\partial (-i\partial_{x_2} )$.

In the following, we give expressions for the scattering wave functions inside a 2D superconductor, a 2D material with Rashba-like spin-orbit coupling for $\vecn =\boldsymbol{x}$ and $\vecn =\boldsymbol{z}$, and a 2D ferromagnet with $\boldsymbol{h}=h\boldsymbol{z}$, treating each material separately. We choose the superconducting region to be located at $x>0$, while the non superconducting region is located at $x<0$.   

\subsubsection{The superconducting region}
\label{s:wfS}

The scattering wave functions on the superconducting side of the interface are
\begin{align}
\begin{split}
    \Psi_n (x_1)=&\Psi_{\text{in},n}^R (x_1) \\
    &+c_{n,1}
    [u_0 \hspace{2.5mm} 0 \hspace{2.5mm} 0 \hspace{2.5mm} v_0]^T e^{iq_x^+ x_1}\\
    &+c_{n,2}[0 \hspace{1mm} -u_0 \hspace{2.5mm} v_0 \hspace{2.5mm} 0]^T e^{iq_x^+ x_1}\\
    &+d_{n,1}[0 \hspace{1mm} -v_0 \hspace{2.5mm} u_0 \hspace{2.5mm} 0]^T e^{-iq_x^- x_1}\\
    &+d_{n,2}[v_0 \hspace{2.5mm} 0 \hspace{2.5mm} 0 \hspace{2.5mm} u_0]^T e^{-iq_x^- x_1},\hspace{2mm}x_1>0,\\
\end{split}\\
\begin{split}
    \Tilde{\Psi}_m (x_2)=&\Tilde{\Psi}_{\text{in},m}^R (x_2) \\
    &+\Tilde{c}_{m,1}[u_0 \hspace{2.5mm} 0 \hspace{2.5mm} 0 \hspace{2.5mm} v_0 ]^T e^{iq_x^+ x_2}\\
    &+\Tilde{c}_{m,2}[0 \hspace{1mm} -u_0 \hspace{2.5mm} v_0 \hspace{2.5mm} 0 ]^T e^{iq_x^+ x_2}\\
    &+\Tilde{d}_{m,1}[0 \hspace{1mm} -v_0 \hspace{2.5mm} u_0 \hspace{2.5mm} 0 ]^T e^{-iq_x^- x_2}\\
    &+\Tilde{d}_{m,2}[v_0 \hspace{2.5mm} 0 \hspace{2.5mm} 0 \hspace{2.5mm} u_0 ]^T e^{-iq_x^- x_2},\hspace{2mm}x_2>0,\\
\end{split}
\end{align}
where the quasi-particles incoming from the right are described by the wave functions 
\begin{equation}
\begin{split}
    &\Psi_{\text{in},5}^R (x_1 )=[u_0 \hspace{2.5mm} 0 \hspace{2.5mm} 0 \hspace{2.5mm} v_0]^T e^{-iq^+_x x_1}\\
    &\Psi_{\text{in},6}^R (x_1 )=[0 \hspace{1mm} -u_0 \hspace{2.5mm} v_0 \hspace{2.5mm} 0]^T e^{-iq^+_x x_1},\\
    &\Psi_{\text{in},7}^R (x_1)= [0 \hspace{1mm} -v_0 \hspace{2.5mm} u_0 \hspace{2.5mm} 0]^T e^{iq^-_x x_1},\\
    &\Psi_{\text{in},8}^R (x_1)=
    [v_0 \hspace{2.5mm} 0 \hspace{2.5mm} 0 \hspace{2.5mm} u_0]^T e^{iq^-_x x_1},\\
\end{split}
\end{equation}
and
\begin{equation}
\begin{split}
    &\Tilde{\Psi}_{\text{in},5}^R (x_2)=[u_0 \hspace{2.5mm} 0 \hspace{2.5mm} 0 \hspace{2.5mm} v_0 ]^T e^{-iq^+_x x_2},\\
    &\Tilde{\Psi}_{\text{in},6}^R (x_2)=[0 \hspace{1mm} -u_0 \hspace{2.5mm} v_0 \hspace{2.5mm} 0 ]^T e^{-iq^+_x x_2},\\
    &\Tilde{\Psi}_{\text{in},7}^R (x_2)=  [0 \hspace{1mm} -v_0 \hspace{2.5mm} u_0 \hspace{2.5mm} 0 ]^T e^{iq^-_x x_2},\\
    &\Tilde{\Psi}_{\text{in},8}^R (x_2 )=[v_0 \hspace{2.5mm} 0 \hspace{2.5mm} 0 \hspace{2.5mm} u_0 ]^T e^{iq^-_x x_2}.\\
\end{split}
\end{equation}
$\Psi_{\text{in},1}^R (x_1)=\Psi_{\text{in},2}^R (x_1)=\Psi_{\text{in},3}^R (x_1)=\Psi_{\text{in},4}^R (x_1 )=\Tilde{\Psi}_{\text{in},1}^R (x_2 )=\Tilde{\Psi}_{\text{in},2}^R (x_2 )=\Tilde{\Psi}_{\text{in},3}^R (x_2 )=\Tilde{\Psi}_{\text{in},4}^R (x_2 )=0$. We reserve the indices $n,m=\{1,2,3,4\}$ for scattering processes with particles or quasi-particles scattering at the interface from the left. 
Above, 
\begin{equation}
    q_x^{\pm}=\{-p_y^2 +\eta[\mu\pm\sqrt{(\omega+i\delta)^2-|\Delta|^2}]\}^{1/2}
\end{equation}
are the allowed $k_x$ values, and
\begin{align}
u_0^2 \equiv \frac{1}{2}[1+\sqrt{(\omega+i\delta) ^2-|\Delta|^2}/(\omega+i\delta)],\\
v_0^2 \equiv \frac{1}{2}[1-\sqrt{(\omega+i\delta)^2-|\Delta|^2}/(\omega+i\delta) ].
\end{align}

\subsubsection{The region with Rashba spin-orbit coupling, $\vecn =\boldsymbol{x}$}

The scattering wave functions on the side of the interface with Rashba-like spin-orbit coupling are
\begin{align}
\begin{split}
    \Psi_n (x_1)=&\Psi_{\text{in},n}^L (x_1)\\
    &+a_{n,1}[1 \hspace{2.5mm} 0 \hspace{2.5mm} 0 \hspace{2.5mm} 0]^T e^{-ik_x^{e,\uparrow}x_1}\\
    &+a_{n,2}[0 \hspace{2.5mm} 1 \hspace{2.5mm} 0 \hspace{2.5mm} 0]^T e^{-ik_x^{e,\downarrow}x_1}\\
    &+b_{n,1}[0 \hspace{2.5mm} 0 \hspace{2.5mm} 1 \hspace{2.5mm} 0]^T e^{ik_x^{h,\uparrow}x_1}\\
    &+b_{n,2}[0 \hspace{2.5mm} 0 \hspace{2.5mm} 0 \hspace{2.5mm} 1]^T e^{ik_x^{h,\downarrow}x_1},\hspace{2mm}x_1<0,\\
\end{split}\label{wf_2Dx1}\\
\begin{split}    
    \Tilde{\Psi}_m (x_2)=&\Tilde{\Psi}_{\text{in},m}^L (x_2 )\\
    &+\Tilde{a}_{m,1}[1 \hspace{2.5mm} 0 \hspace{2.5mm} 0 \hspace{2.5mm} 0]^T e^{-ik_x^{e,\uparrow}x_2}\\
    &+\Tilde{a}_{m,2}[0 \hspace{2.5mm} 1 \hspace{2.5mm} 0 \hspace{2.5mm} 0]^T e^{-ik_x^{e,\downarrow}x_2}\\
    &+\Tilde{b}_{m,1}[0 \hspace{2.5mm} 0 \hspace{2.5mm} 1 \hspace{2.5mm} 0]^T e^{ik_x^{h,\uparrow}x_2}\\
    &+\Tilde{b}_{m,2}[0 \hspace{2.5mm} 0 \hspace{2.5mm} 0 \hspace{2.5mm} 1]^T e^{ik_x^{h,\downarrow}x_2},\hspace{2mm}x_2<0,\\
\end{split}
\label{wf_2Dx2}
\end{align}
if $\vecn =\boldsymbol{x}$. The particles incoming from the left are described by
\begin{equation}
\begin{split}
    &\Psi_{\text{in},1}^L (x_1 )=[1 \hspace{2.5mm} 0 \hspace{2.5mm} 0 \hspace{2.5mm} 0]^T e^{ik_x^{e,\uparrow}x_1},\\
    &\Psi_{\text{in},2}^L (x_1 )=[0 \hspace{2.5mm} 1 \hspace{2.5mm} 0 \hspace{2.5mm} 0]^T e^{ik_x^{e,\downarrow}x_1},\\
    &\Psi_{\text{in},3}^L (x_1 )=[0 \hspace{2.5mm} 0 \hspace{2.5mm} 1 \hspace{2.5mm} 0]^T e^{-ik_x^{h,\uparrow}x_1},\\
    &\Psi_{\text{in},4}^L (x_1)=[0 \hspace{2.5mm} 0 \hspace{2.5mm} 0 \hspace{2.5mm} 1]^T e^{-ik_x^{h,\downarrow}x_1},\\
\end{split}
\label{x_incoming}
\end{equation}
and
\begin{equation}
\begin{split}
    &\Tilde{\Psi}_{\text{in},1}^L (x_2 )=[1 \hspace{2.5mm} 0 \hspace{2.5mm} 0 \hspace{2.5mm} 0]^T e^{ik_x^{e,\uparrow}x_2},\\
    &\Tilde{\Psi}_{\text{in},2}^L (x_2 )=[0 \hspace{2.5mm} 1 \hspace{2.5mm} 0 \hspace{2.5mm} 0]^T e^{ik_x^{e,\downarrow}x_2},\\
    &\Tilde{\Psi}_{\text{in},3}^L (x_2 )=[0 \hspace{2.5mm} 0 \hspace{2.5mm} 1 \hspace{2.5mm} 0]^T e^{-ik_x^{h,\uparrow}x_2},\\
    &\Tilde{\Psi}_{\text{in},4}^L (x_2 )=[0 \hspace{2.5mm} 0 \hspace{2.5mm} 0 \hspace{2.5mm} 1]^T e^{-ik_x^{h,\downarrow}x_2}.\\
\end{split}
\label{x_wfTilde}
\end{equation}
$\Psi_{\text{in},5}^L (x_1)=\Psi_{\text{in},6}^L (x_1 )=\Psi_{\text{in},7}^L (x_1 )=\Psi_{\text{in},8}^L (x_1 )=\Tilde{\Psi}_{\text{in},5}^L (x_2 )=\Tilde{\Psi}_{\text{in},6}^L (x_2)=\Tilde{\Psi}_{\text{in},7}^L (x_2)=\Tilde{\Psi}_{\text{in},8}^L (x_2)=0$. We reserve the indices $n,m=\{5,6,7,8\}$ for scattering processes with particles incoming from the right. Above,
\begin{equation}
    k_x^{e(h),\uparrow(\downarrow)}=\{-p_y^2 +\eta[\mu\pm(\omega+i\delta \pm'\lambda p_y )]\}^{1/2}
\end{equation}
are the allowed $k_x$ values.  $\pm$ correspond to electrons and holes, respectively, while $\pm'$ correspond to spin up and spin down, respectively.

\subsubsection{The ferromagnetic scattering wave functions}

The scattering wave functions for a ferromagnet with $\boldsymbol{h}=h\boldsymbol{z}$, are on the same form as for a material with Rashba spin-orbit coupling where $\vecn=\boldsymbol{x}$, and are thus given by Eqs.~\eqref{wf_2Dx1}, \eqref{wf_2Dx2}, \eqref{x_incoming}, and \eqref{x_wfTilde}. 
The allowed $k_x$ values are in this case given by 
\begin{equation}
    k_x^{e(h),\uparrow(\downarrow)}= \{-p_y^2 +\eta[\mu\pm(\omega +i\delta)\mp' h]\}^{1/2}.
\end{equation}
$\pm$ refers to electrons and holes, respectively, and $\mp'$ refers to spin up and spin down, respectively.

\subsubsection{The region with Rashba spin-orbit coupling, $\vecn =\boldsymbol{z}$}
\label{s:wavefunctions2Dz}

The scattering wave functions on the side of the interface with Rashba-like spin-orbit coupling are
\begin{align}
\begin{split}
    \Psi_n (x_1)&=\Psi_{\text{in},n}^L (x_1 )
    \\
    &+a_{n,1}
    [\phantom{-}1 \hspace{2.5mm} ie^{i\phi} \hspace{2.5mm} 0 \hspace{2.5mm} 0\hspace{1.8mm}]^T e^{-ik_x^{e,+}x_1}\\
    &+a_{n,2}
    [-1 \hspace{2.5mm} ie^{i\phi} \hspace{2.5mm} 0 \hspace{2.5mm} 0\hspace{1.8mm}]^T e^{-ik_x^{e,-}x_1}\\
    &+b_{n,1}
    [0 \hspace{2.5mm} 0 \hspace{2.5mm} \phantom{-}1 \hspace{2.5mm} ie^{-i\phi}]^T e^{ik_x^{h,-}x_1}\\
    &+b_{n,2}
    [0 \hspace{2.5mm} 0 \hspace{1mm} -1 \hspace{2.5mm} ie^{-i\phi}]^T e^{ik_x^{h,+}x_1},\hspace{2mm}x_1<0,\\
\end{split}\\
\begin{split}
    \Tilde{\Psi}_m (x_2)&=\Tilde{\Psi}_{\text{in},m}^L (x_2)\\
    &+\Tilde{a}_{m,1}
    [\phantom{-}1 \hspace{2.5mm} ie^{i\phi} \hspace{2.5mm} 0 \hspace{2.5mm} 0\hspace{1.8mm}]^T e^{-ik_x^{e,+}x_2}\\
    &+\Tilde{a}_{m,2}
    [-1 \hspace{2.5mm} ie^{i\phi} \hspace{2.5mm} 0 \hspace{2.5mm} 0\hspace{1.8mm}]^T e^{-ik_x^{e,-}x_2}\\
    &+\Tilde{b}_{m,1}
    [0 \hspace{2.5mm} 0 \hspace{2.5mm} \phantom{-}1 \hspace{2.5mm} ie^{-i\phi}]^T e^{ik_x^{h,-}x_2}\\
    &+\Tilde{b}_{m,2}
    [0 \hspace{2.5mm} 0 \hspace{1mm} -1 \hspace{2.5mm} ie^{-i\phi}]^T e^{ik_x^{h,+}x_2},\hspace{2mm}x_2<0,\\
\end{split}
\end{align}
if $\vecn =\boldsymbol{z}$. The particles incoming from the left are described by
\begin{equation}
\begin{split}
    &\Psi_{\text{in},1}^L (x_1)=[\phantom{-}1 \hspace{2.5mm} ie^{i\phi} \hspace{2.5mm} 0 \hspace{2.5mm} 0\hspace{1.8mm}]^T e^{ik_x^{e,+}x_1},\\
    &\Psi_{\text{in},2}^L (x_1 )=[-1 \hspace{2.5mm} ie^{i\phi} \hspace{2.5mm} 0 \hspace{2.5mm} 0\hspace{1.8mm}]^T e^{ik_x^{e,-}x_1},\\
    &\Psi_{\text{in},3}^L (x_1 )=[0 \hspace{2.5mm} 0 \hspace{2.5mm} \phantom{-}1 \hspace{2.5mm} ie^{-i\phi}]^T e^{-ik_x^{h,-}x_1},\\
    &\Psi_{\text{in},4}^L (x_1)=[0 \hspace{2.5mm} 0 \hspace{1mm} -1 \hspace{2.5mm} ie^{-i\phi}]^T e^{-ik_x^{h,+}x_1}\\
\end{split}
\end{equation}
and
\begin{equation}
\begin{split}
    &\Tilde{\Psi}_{\text{in},1}^L (x_2 )=[\phantom{-}1 \hspace{2.5mm} ie^{i\phi} \hspace{2.5mm} 0 \hspace{2.5mm} 0\hspace{1.8mm}]^T e^{ik_x^{e,+}x_2},\\
    &\Tilde{\Psi}_{\text{in},2}^L (x_2 )=[-1 \hspace{2.5mm} ie^{i\phi} \hspace{2.5mm} 0 \hspace{2.5mm} 0\hspace{1.8mm}]^T e^{ik_x^{e,-}x_2},\\
    &\Tilde{\Psi}_{\text{in},3}^L (x_2 )=[0 \hspace{2.5mm} 0 \hspace{2.5mm} \phantom{-}1 \hspace{2.5mm} ie^{-i\phi}]^T e^{-ik_x^{h,-}x_2},\\
    &\Tilde{\Psi}_{\text{in},4}^L (x_2 )=[0 \hspace{2.5mm} 0 \hspace{1mm} -1 \hspace{2.5mm} ie^{-i\phi}]^T e^{-ik_x^{h,+}x_2}.\\
\end{split}
\end{equation}
$\Psi_{\text{in},5}^L (x_1)=\Psi_{\text{in},6}^L (x_1 )=\Psi_{\text{in},7}^L (x_1 )=\Psi_{\text{in},8}^L (x_1 )=\Tilde{\Psi}_{\text{in},5}^L (x_2 )=\Tilde{\Psi}_{\text{in},6}^L (x_2)=\Tilde{\Psi}_{\text{in},7}^L (x_2 )=\Tilde{\Psi}_{\text{in},8}^L (x_2)=0$. Above,
$k_x^{e(h),+(-)} = k^{e(h),+(-)}\cos(\phi)$ are the allowed $k_x$ values, where
\begin{equation}
    k^{e(h),+(-)}=\{[(\lambda\eta/2)^2 +\eta(\mu\pm \omega +i\delta)]^{1/2} \pm\pm'\lambda\eta/2\}^{1/2}.
\end{equation}
$\pm$ correspond to electrons and holes, respectively, and $\pm'$ correspond to the two different spin-mixed states. We define $k_x^{e(h),+(-)}$ to be positive by setting $\phi\in[-\pi/2,\pi/2]$.

\subsection{The singlet and triplet retarded anomalous Green's functions}

From the scattering wave functions, we construct the retarded Green's function in Nambu~$\otimes$~spin space  \cite{cayao_prb_17,cayao_prb_18},
\begin{align}
\begin{split}
    G^r (x_1 & >x_2 ,p_y ;\omega )=\\
    &\sum_{n,m=1}^4 \alpha_{nm}\Psi_{n} (x_1 ,p_y )\Tilde{\Psi}_{m+4}^T (x_2 ,p_y ),\\
\end{split}\label{Gr1}\\
\begin{split}
    G^r (x_1 & <x_2 ,p_y ;\omega )=\\
    &\sum_{n,m=1}^4 \beta_{nm}\Psi_{n+4} (x_1 ,p_y )\Tilde{\Psi}_{m}^T (x_2 ,p_y ).\\
    \end{split}
    \label{Gr2}
\end{align}
The coefficients $\alpha_{nm}$ and $\beta_{nm}$ are found from the boundary conditions of the retarded Green's function at $x_1 =x_2$ \cite{cayao_prb_17,cayao_prb_18}, 
\begin{equation}
    \begin{split}
        &[G^r (x_1 >x_2 ,p_y  ;\omega)]_{x_1 =x_2 }=[G^r (x_1 <x_2 ,p_y  ;\omega)]_{x_1 =x_2 },\\
        &[\partial_{x_1} G^r (x_1 >x_2 ,p_y ;\omega)]_{x_1 =x_2}\\
        &-[\partial_{x_1} G^r (x_1 <x_2 ,p_y ;\omega)]_{x_1 =x_2}=\eta\hat{\tau}_3 \hat{\sigma}_0 ,\\
    \end{split}
\end{equation}
We rewrite to the center of mass coordinate $X\equiv(x_1 +x_2 )/2$, and the relative coordinate $x\equiv x_1 -x_2$, and calculate the even- and odd-frequency singlet and triplet anomalous contributions to the retarded Green's functions according to Eqs.~(4) and (5) in the letter. We use that
\begin{equation}
    \begin{split}
        F_0^{r} (X ,x ,p_y ;\omega)\equiv&[G_{1,4}^{r} (X ,x ,p_y ;\omega)\\
        -&\phantom{[}G^{r}_{2,3} (X ,x ,p_y ;\omega)]/2,\\
        F_1^{r} (X ,x ,p_y  ;\omega)\equiv&\phantom{[}G_{1,3}^{r} (X ,x ,p_y ;\omega),\\
        F_2^{r} (X ,x ,p_y ;\omega)\equiv&\phantom{[}G_{2,4}^{r} (X ,x ,p_y  ;\omega),\\
        F_3^{r} (X ,x ,p_y ;\omega)\equiv&[G_{1,4}^{r} (X ,x ,p_y ;\omega)\\
        +&\phantom{[}G_{2,3}^{r} (X ,x ,p_y ;\omega)]/2.\\
    \end{split}
\end{equation}

\subsubsection{The SOC/S system, $\vecn =\boldsymbol{x}$}

For simplicity of notation, we define $k_{e1}\equiv k_x^{e,\uparrow}$, $k_{e2}\equiv k_x^{e,\downarrow}$, $k_{h1}\equiv k_x^{h,\downarrow}$, and $k_{h2}\equiv k_x^{h,\uparrow}$. On the side of the interface with Rashba-like spin-orbit coupling, where $x_1 ,x_2 <0$, the nonzero even- and odd-frequency singlet and triplet retarded anomalous Green's functions are given by
\begin{equation}
    \begin{split}
        F^{r,E}_0 &(X ,x ,p_y ;\omega)=\frac{\eta}{2i}u_0 v_0 (q_x^+ + q_x^- )\\
        &\sum_{l=1,2} \frac{1}{D_l}e^{-i(k_{el}-k_{hl})X}\cos[(k_{el}+k_{hl})x/2],\\
        F^{r,O}_0 &(X ,x ,p_y ;\omega)=\frac{\eta}{2}u_0 v_0 (q_x^+ + q_x^- )\\
        &\sum_{l=1,2}\frac{1}{D_l}e^{-i(k_{el}-k_{hl})X}\sin[(k_{el}+k_{hl})x/2],\\
        F^{r,E}_3 &(X ,x ,p_y ;\omega)=\frac{\eta}{2i}u_0 v_0 (q_x^+ + q_x^- )\\
        &\sum_{l=1,2}(-1)^{l-1}\frac{1}{D_l}e^{-i(k_{el}-k_{hl})X}\cos[(k_{el}+k_{hl})x/2],\\
        F^{r,O}_3 &(X ,x ,p_y ;\omega)=\frac{\eta}{2}u_0 v_0 (q_x^+ + q_x^- )\\
        &\sum_{l=1,2}(-1)^{l-1}\frac{1}{D_l}e^{-i(k_{el}-k_{hl})X}\sin[(k_{el}+k_{hl})x/2],\\
    \end{split}
    \label{f0f3_analyticalExpressions}
\end{equation}
where 
\begin{equation}
\begin{split}
    D_{l}\equiv& u_0^2 (k_{el}+q_x^+ )(k_{hl}+q_x^- )
    +v_0^2 (k_{hl}-q_x^+ )(-k_{el}+q_x^- ).
\end{split}
\end{equation}
On the superconducting side of the interface, where $x_1 ,x_2 >0$, the nonzero even- and odd-frequency singlet and triplet retarded anomalous Green's functions are given by
\begin{equation}
    \begin{split}
        F_0^{r,E}&(X ,x ,p_y ;\omega )=-\frac{\eta}{2i}\frac{u_0 v_0}{(u_0^2 -v_0^2)}
        \bigg\{e^{i(q_x^+ -q_x^- )X}\\
        &\cos[(q_x^+ +q_x^- )x/2]\sum_{l=1,2}\frac{1}{D_l}(k_{el}+k_{hl})\\
        -&\bigg(\frac{1}{q_x^+}e^{iq_x^+ |x|}+\frac{1}{q_x^-}e^{-iq_x^- |x|}\bigg)\\
        -&\frac{1}{2}\sum_{l=1,2}\bigg(\frac{E_l}{D_l}\frac{1}{q_x^+}e^{2iq_x^+ X}
        +\frac{F_l}{D_l}\frac{1}{q_x^-}e^{-2iq_x^- X}\bigg)\bigg\},\\
        F_0^{r,O}&(X ,x ,p_y ;\omega )=\frac{\eta}{2}u_0 v_0 e^{i(q_x^+ -q_x^- )X}\\
        &\sin[(q_x^+ +q_x^- )x/2]
        \sum_{l=1,2}\frac{1}{D_l}(k_{el}+k_{hl}),\\
        F_3^{r,E}&(X ,x ,p_y ;\omega )=-\frac{\eta}{2i}\frac{u_0 v_0}{(u_0^2 -v_0^2)}
        \bigg\{e^{i(q_x^+ -q_x^- )X}\\
        &\cos[(q_x^+ +q_x^- )x/2]
        \sum_{l=1,2}(-1)^{l-1}\frac{1}{D_l}(k_{el}+k_{hl})\\
        -&\frac{1}{2}\sum_{l=1,2}(-1)^{l-1}\bigg(\frac{E_l}{D_l}\frac{1}{q_x^+}e^{2iq_x^+ X}
        +\frac{F_l}{D_l}\frac{1}{q_x^-}e^{-2iq_x^- X}\bigg)\bigg\},\\
        F_3^{r,O}&(X ,x ,p_y ;\omega )=\frac{\eta}{2}u_0 v_0
        e^{i(q_x^+ -q_x^- )X}\\
        &\sin[(q_x^+ +q_x^- )x/2]
        \sum_{l=1,2}(-1)^{l-1}\frac{1}{D_l}(k_{el}+k_{hl}),\\
    \end{split}
    \label{GF_nx_insideSC}
\end{equation}
where
\begin{equation}
    \begin{split}
        E_{l}\equiv& u_0^2 (k_{el}-q_x^+ )(-k_{hl} -q_x^- )
        +v_0^2 (k_{hl}+q_x^+ )(k_{el}-q_x^- ),\\
        F_{l}\equiv& u_0^2 (k_{el}+q_x^+ )(-k_{hl} +q_x^- )
        +v_0^2 (k_{hl}-q_x^+ )(k_{el}+q_x^- ).
    \end{split}
\end{equation}
There are no equal-spin triplets in the system. 

\subsubsection{The F/S system}

The even- and odd-frequency singlet and triplet retarded anomalous Green's functions of the F/S system are given by the same expressions as for the SOC/S system with $\vecn=\boldsymbol{x}$ if we let $F_3^{r,E}(X,x,p_y ;\omega)\leftrightarrow F_3^{r,O}(X,x,p_y ;\omega)$ in Eqs.~\eqref{f0f3_analyticalExpressions} and \eqref{GF_nx_insideSC}.
There are no equal-spin triplets in the system.

\subsection{The symmetries of the singlet and triplet retarded anomalous Green's functions}

Finally, we investigate the spatial symmetries of the singlet and triplet retarded anomalous Green's functions of the SOC/S systems with $\vecn =\boldsymbol{x}$ and $\vecn =\boldsymbol{z}$ and the F/S system with $\vecm||\boldsymbol{z}$. P$_x$ is inversion of the relative $x$ coordinate, $x\to-x$. P$_y$ is inversion of the momentum along the $y$ axis, $p_y \to -p_y$. $P$ is total spatial inversion, and must be 1 for $F_0^E$ and $F_3^O$, and -1 for $F_0^O$ and $F_3^E$, according to the Pauli principle. For $\text{P}=1$, we may have $\text{P}_x =\text{P}_y =1$, which describes an $s$- or a $d_{x^2-y^2}$-wave amplitude, or $\text{P}_x = \text{P}_y =-1$, which describes a $d_{xy}$-wave amplitude. For $\text{P} =-1$, we may have $\text{P}_x =1$ and $\text{P}_y =-1$, which describes a $p_y$-wave amplitude, or $\text{P}_x =-1$ and $\text{P}_y =1$, which describes a $p_x$-wave amplitude. 
Considering P, $\text{P}_x$, and $\text{P}_y$ is not sufficient for determining whether a Green's function has an $s$-wave or a $d_{x^2 -y^2}$-wave symmetry. In order to prove the presence of $s$-wave singlets and triplets, we apply the Fourier transform, 
\begin{equation}
\begin{split}
        &F^{r,E(O)}_{0(3)}(X,p_x ,p_y ;\omega)\\
        &=\int_{\infty}^{\infty}dx\hspace{1mm}F^{r,E(O)}_{0(3)}(X,x ,p_y ;\omega)e^{-i p_x x},
\end{split}
\end{equation}
and set $p_x$ and $p_y$ to zero, which is equivalent to integrating over all spatial coordinates.

\subsubsection{The SOC/S system, $\vecn =\boldsymbol{x}$}

\begin{table}[h]
    \centering
    \begin{tabular}{cccc}
    \hline
              & P$_x$   & P$_y$ & P\\
    \hline
    $F_0^{r,E}$     & $\phantom{-}1$ & $\phantom{-}1$ & $\phantom{-}1$  \\
    $F_0^{r,O}$    & $-1$ & $\phantom{-}1$ & $-1$  \\
    $F_3^{r,E}$     & $\phantom{-}1$ & $-1$ & $-1$  \\
    $F_3^{r,O}$     & $-1$ & $-1$ & $\phantom{-}1$  \\
    \hline
    \end{tabular}
    \caption{The above table shows the parities of the SOC/S system with $\vecn=\boldsymbol{x}$ under $x\to -x$ (P$_x$), $p_y \to -p_y$ (P$_y$), and total spatial inversion (P) for the nonzero singlet and triplet even- and odd-frequency retarded anomalous Green's functions given in Eqs.~\eqref{f0f3_analyticalExpressions} and \eqref{GF_nx_insideSC}.}
    \label{tab:PxPy}
\end{table}
The symmetries of the Green's functions in Eqs.~\eqref{f0f3_analyticalExpressions} and \eqref{GF_nx_insideSC} under P$_x$, P$_y$ and P are given in Table \ref{tab:PxPy}. We see from the table that $F_0^{r,E}$ can represent $s$- and $d_{x^2 -y^2}$-wave singlets, $F_0^{r,O}$ represents $p_x$-wave singlets, $F_3^{r,E}$ represents a $p_y$-wave opposite-spin triplets, and $F_3^{r,O}$ represents $d_{xy}$-wave opposite-spin triplets. By integrating over all of space, we find that $s$-wave singlets are present.

\subsubsection{The SOC/S system, $\vecn =\boldsymbol{z}$}

The symmetries of the Green's functions of the SOC/S system for $\vecn =\boldsymbol{z}$ are shown in Table \ref{tab:PxPy_nz}. These were found numerically by the same approach as for the two other systems.
We see that the same singlet and opposite-spin triplet amplitudes are present as for $\vecn =\boldsymbol{x}$. In addition, we have nonzero equal-spin triplet amplitudes, that are a mix of triplet amplitudes with different symmetries under P$_x$ and P$_y$. $F_1^{r,E}$ and $F_2^{r,E}$ are therefore a mix of $p_x$- and $p_y$-wave even-frequency triplets, while $F_1^{r,O}$ and $F_2^{r,O}$ are a mix $s$- and $d$-wave triplets. 
\begin{table}[h]
    \centering
    \begin{tabular}{cccc}
    \hline
              & P$_x$   & P$_y$ & P\\
    \hline
    $F_0^{r,E}$     & $\phantom{-}1$ & $\phantom{-}1$ & $\phantom{-}1$  \\
    $F_0^{r,O}$    & $-1$ & $\phantom{-}1$ & $-1$  \\
    $F_3^{r,E}$     & $\phantom{-}1$ & $-1$ & $-1$  \\
    $F_3^{r,O}$     & $-1$ & $-1$ & $\phantom{-}1$  \\
    $F_1^{r,E}$ & - & - & $-1$\\
    $F_1^{r,O}$&-&-&$\phantom{-}1$\\
    $F_2^{r,E}$&-&-&$-1$\\
    $F_2^{r,O}$&-&-&$\phantom{-}1$\\
    \hline
    \end{tabular}
    \caption{The above table shows the parities of the SOC/S system with $\vecn=\boldsymbol{z}$ under $x_1 \leftrightarrow -x_2$ (P$_x$), $p_y \to -p_y $ (P$_y$), and total spatial inversion (P) for the singlet and triplet even- and odd-frequency retarded anomalous Green's functions present in the system. In addition to the singlets and triplets present for $\vecn =\boldsymbol{x}$ shown in Table \ref{tab:PxPy}, we have nonzero equal-spin triplet amplitudes with mixing (-) of the possible symmetries in P$_x$ and P$_y$.}
    \label{tab:PxPy_nz}
\end{table}

\subsubsection{The F/S system}

Table \ref{tab:PxPy_FM} shows the symmetries of the even- and odd-frequency singlet and triplet retarded anomalous Green's functions of the F/S system under P$_x$, P$_y$, and P. Due to the lack of symmetry-breaking in the $y$ direction, we must conclude that there are no $p_y$- or $d$-wave symmetries present.
The nonzero singlet and triplet retarded anomalous Green's functions are therefore the $s$-wave even-frequency singlet, the $p_x$-wave odd-frequency singlet, the $p_x$-wave even-frequency opposite-spin triplet and the $s$-wave odd-frequency opposite-spin triplet. 
\begin{table}[h]
    \centering
    \begin{tabular}{cccc}
    \hline
              & P$_x$   & P$_y$ & P\\
    \hline
    $F_0^{r,E}$     & $\phantom{-}1$ & $\phantom{-}1$ & $\phantom{-}1$  \\
    $F_0^{r,O}$    & $-1$ & $\phantom{-}1$ & $-1$  \\
    $F_3^{r,E}$     & $-1$ & $\phantom{-}1$ & $-1$  \\
    $F_3^{r,O}$     & $\phantom{-}1$ & $\phantom{-}1$ & $\phantom{-}1$  \\
    \hline
    \end{tabular}
    \caption{The above table shows the parities of the F/S system under $x\to -x$ (P$_x$), $p_y \to -p_y$ (P$_y$), and total spatial inversion (P).}
    \label{tab:PxPy_FM}
\end{table}

%

\end{document}